\documentclass[twocolumn,nofootinbib,superscriptaddress,notitlepage,preprintnumbers]{revtex4-1}

\usepackage{graphicx}
\usepackage{dcolumn}
\usepackage{bm}
\usepackage{hyperref}
\usepackage{bigints}
\usepackage{multirow}
\usepackage{enumitem}
\usepackage{upgreek}
\usepackage{url}
\usepackage{amsmath}
\usepackage{amssymb}

\usepackage[english]{babel}
\usepackage{xcolor}
\hypersetup{
    colorlinks,
    linkcolor={red!80!black},
    citecolor={blue!50!black},
    urlcolor={blue!90!black}
}

\usepackage{slashed,cancel}
\usepackage{xspace}

\usepackage[utf8]{inputenc}
	
\usepackage[T2A]{fontenc}
\usepackage[utf8]{inputenc}														
\usepackage{mathtext}

\usepackage{amsmath,amsfonts,amssymb,mathtools}	
\usepackage{icomma} 

\newcommand{\pc}[1]{#1}
\newcommand{\jb}[1]{#1}

\newcommand{\lm}[1]{#1}
\usepackage{babel,csquotes,xpatch}

\begin{document}

\preprint{IFIC/23-12}
\preprint{IFT-UAM/CSIC-23-35}
\preprint{FTUV-23-0329.0644}

\title{New Physics searches using ProtoDUNE and the CERN SPS accelerator}

\newcommand{\IFT}
{Instituto de Física Teórica UAM-CSIC, Calle Nicolás Cabrera 13-15, Cantoblanco E-28049 Madrid, Spain}

\newcommand{\IFIC}
{Instituto de Física Corpuscular, Universidad de Valencia and CSIC, Carrer del Catedrátic José Beltrán Martinez, 2, 46980 Paterna, Valencia, Spain
}

\author{Pilar Coloma}
\email[]{pilar.coloma@ift.csic.es} 
\affiliation{\IFT}

\author{Jacobo L\'opez-Pav\'on}
\email[]{jlpavon@ific.uv.es} 
\affiliation{\IFIC}

\author{Laura Molina-Bueno}
\email[]{laura.molina.bueno@cern.ch}
\affiliation{\IFIC}

\author{Salvador Urrea}
\email[]{salvador.urrea@ific.uv.es}
\affiliation{\IFIC}

\begin{abstract}
    The exquisite capabilities of liquid Argon Time Projection Chambers make them ideal to search for weakly interacting particles in Beyond the Standard Model scenarios.
    Given their location at CERN the ProtoDUNE detectors may be exposed to a flux of such particles, produced in the collisions of 400~GeV protons (extracted from the Super Proton Synchrotron accelerator) on a target. Here we point out the interesting possibilities that such a setup offers to search for both long-lived unstable particles (Heavy Neutral Leptons, axion-like particles, etc) and stable particles (e.g. light dark matter, or millicharged particles). Our results show that, under conservative assumptions regarding the expected luminosity, this setup has the potential to improve over present bounds for some of the scenarios considered. This could be done within a short timescale, using facilities that are already in place at CERN, and without interfering with the experimental program in the North Area at CERN.
\end{abstract}

\date{\today}

\maketitle

\section{Introduction}

Despite the intensive searches in the last decades, no conclusive signal of physics Beyond the Standard Model (BSM) has been observed at the Large Hadron Collider or at dark matter (DM) direct detection experiments. A possible explanation for this is that the new physics (NP) is weakly coupled to the visible sector and lies at low scales. Models of this sort could easily evade current searches, which have so far been aimed at new particles with masses at or above the electroweak scale. From the theory side, several low-scale new physics scenarios have been recently put forward as an alternative approach to address some of the most pressing questions in particle physics: the origin of neutrino masses, the observed baryon asymmetry of the universe, and the DM problem. Weakly interacting particles arise naturally in these type of scenarios, as it is for instance the case of low-scale seesaw models, which are able to answer the first two. Regarding the DM problem, an exciting possibility is based on the existence of an extended ``dark'' (or hidden) sector communicating with the Standard Model (SM) via a new (light) mediator that is weakly coupled. Different types of particles can act as mediators between the two sectors, covering a wide range of masses, and leading to distinct phenomenological consequences.

This change of paradigm has boosted a worldwide effort yielding a plethora of novel approaches and proposals, in particular, for experiments lying at the edge of the intensity frontier (see for instance~\cite{Agrawal:2021dbo}). In this sense, one of the most competitive searches are those performed at {\it beam-dump} experiments, where the collision of high-energy particles (typically protons, or electrons) against a target may produce an intense flux of new light states, typically from meson decays. Being weakly interacting, such particles could propagate over long distances before decaying visibly (or interacting) inside a detector placed downstream. An example of such facilities are neutrino experiments, where the collision of high-intensity proton beams sourcing the neutrino flux, can also be used to produce a variety of new particles which may lead to observable signals in neutrino detectors. 

In this Letter, we propose a new beam-dump experiment using the existing ProtoDUNE detectors at the CERN Neutrino Platform, two kiloton-scale liquid Argon Time Projection Chambers (LArTPCs) constructed to prototype and consolidate the technology of the DUNE Far Detector~\cite{DUNE:2017pqt, DUNE:2020lwj}. These detectors are downstream with respect to the CERN's North Area targets, used to produce secondary charged particle beams from the interactions of protons extracted from the CERN Super Proton Synchrotron (SPS) accelerator. As a result, the proton collisions in the primary target may generate a flux of BSM particles which could leave a visible signal in the ProtoDUNE detectors. To the best of our knowledge, this is the first time that this has been pointed out in the literature. Most importantly, we stress out that searches for such signals can be carried out parasitically, without interfering with the dense experimental program at the CERN North Area Experimental Hall (EHN1) multipurpose facility.  

One of the key features of the LArTPCs is their excellent imaging capabilities, which allow them to fully reconstruct the tracks of ionising particles resulting from the decay or scattering of the produced new particles. This feature, together with the potential time synchronisation with the beam, can significantly reduce the possible background sources. This first advantage is crucial for surface-detectors exposed to a huge flux of cosmics such as ProtoDUNE. The second intrinsic advantage that this proposal offers is the wider phase space that can be covered, compared to similar searches at neutrino experiments such as  T2K~\cite{T2K:2019jwa} or MicroBooNE~\cite{MicroBooNE:2019izn} (for future prospects see Ref.~\cite{Abdullahi:2022jlv}), thanks to the higher proton beam energy available at the SPS (400~GeV, as opposed to the $80-120$~GeV protons foreseen for example at DUNE~\cite{DUNE:2020ypp}). This allows not only to abundantly produce light short-lived mesons (such as $\pi^0, \eta, \eta'$, etc) but also to produce a significant flux of heavier short-lived meson (such as $D$, $D_s$, $B$, or $\Upsilon$), as in the SHIP~\cite{Aberle:2839677} or SHADOWS~\cite{Alviggi:2839484} experiments. The beam configuration, without a decay volume, does not allow to study decays from longer-lived mesons such as charged kaons or pions since they are significantly deviated with a set of magnets located after the primary target. Nevertheless, this peculiarity translates in an intrinsic advantage as the background from SM neutrinos will be significantly reduced in this case, as opposed to neutrino experiments. Finally, thanks to both its large volume and the high density of liquid Argon (LAr), ProtoDUNE can be used to search for \emph{both} unstable and stable weakly interacting particles produced in this manner. \pc{In particular, we highlight here the advantages offered by LAr TPC detectors thanks to their very low detection thresholds (for a recent review see e.g., Ref.~\cite{Caratelli:2022llt} and references therein). } We will therefore consider both scenarios separately in this work. For each of them, we will first consider a model-independent approach in such a way that our results can be recasted to specific scenarios, addressing then the case of two well-motivated specific models.

The rest of the paper is structured as follows. We provide an overview of the experimental setup in Sec.~\ref{sec:setup}. Our results are presented in Sec.~\ref{sec:results}, where the cases of long-lived states and stable particles are discussed separately. Finally, we summarize and conclude in Sec.~\ref{sec:summary}. 

\section{Experimental setup}
\label{sec:setup}

The two ProtoDUNE detectors were constructed and installed in the CERN Neutrino Platform, approved experiments NP02 (ProtoDUNE-VD) and NP04 (ProtoDUNE-SP/ProtoDUNE-HD), at the end of EHN1~\cite{Charitonidis:2017omo,DUNE:2020cqd}. 
To produce the secondary beams in the North Area at CERN, the high-energy, high-intensity proton beam extracted from the SPS accelerator impinges on a thin (50~cm) Beryllium target, T2. Downstream the target, secondary particles are selected with the use of magnetic spectrometers and transported to the various experimental areas. In particular, the ProtoDUNE detectors are relatively aligned with the secondary H2/H4 beamlines and thus with the primary target T2. This feature puts them in a unique position to act as a beam-dump experiment, being located at a distance of $L\sim 677$ and $723$~m from the target.

\begin{center}
    \begin{figure*}[htb!]
        \includegraphics[width=0.95\textwidth]{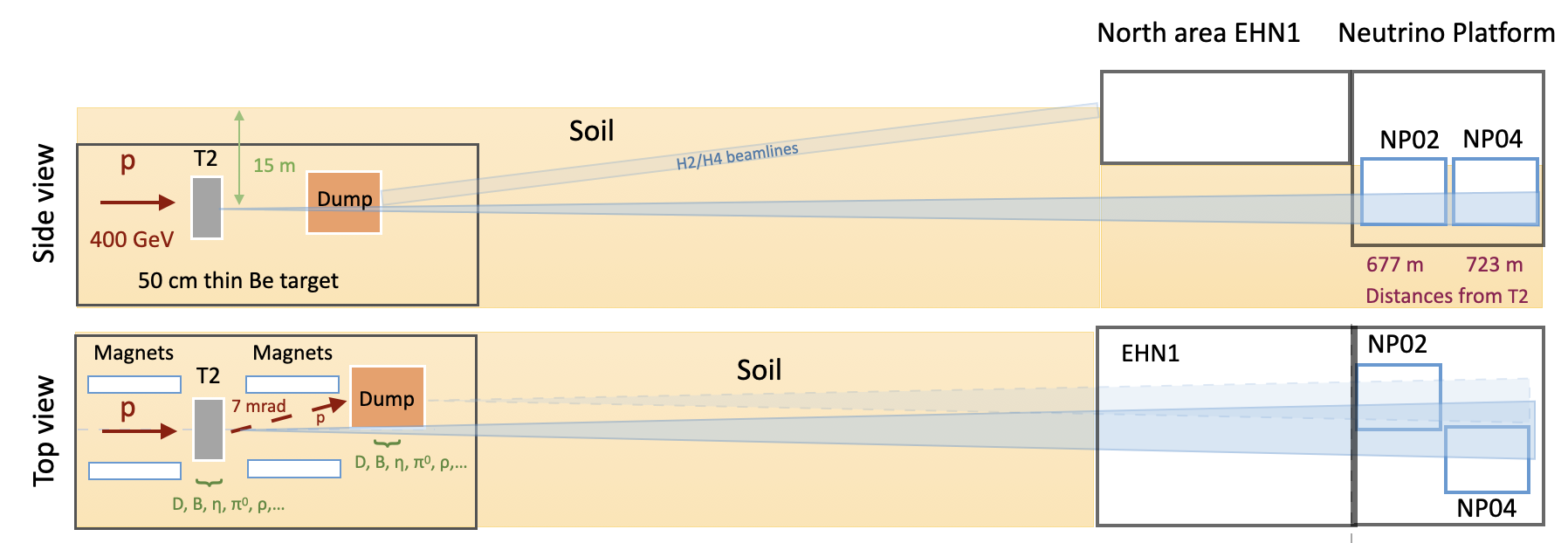}
        \caption{Sketch of the experimental configuration considered in this study. The upper panel (side view) indicates the location of the EHN1 Hall, where the ProtoDUNE modules are installed, with respect to the T2 target and the beam dump (TAX). The target is $\sim 15$~m underground, and there are $\sim 500$~m of soil between the beam dump and EHN1. In the lower panel (top view), the  cones indicate the direction of the beam of BSM particles: solid for the particles produced in the target, and dashed for those produced in the dump. The direction of the proton beam is indicated by the red arrows, see text for details.
        \label{fig:setup}}
    \end{figure*}
\end{center}

A sketch of the experimental configuration considered in this study is illustrated in Fig.~\ref{fig:setup}. About $2-7\times 10^{12}$ protons per spill are extracted from the SPS accelerator to the target,
with a spill duration of 4.8 s. \lm{The repetition rate varies between 14.4 and 60 seconds depending on the experiments needs, with 3000 spills/day on average}. Therefore, in a year $\sim 3.5\times10^{18}$ protons on target (POT) are dumped against T2~\footnote{
The duty cycle considered for our study has been set to that of SPS during 2022 (see~\cite{SPSweb}). Specifically, for a data-taking period of 6 months and with the beam parameters listed above, this leads to a duty cycle of $30\%$.}. The T2 target is located 15~m underground for radiation shielding and is surrounded by a series of magnets, collimators and other beamline elements depending on the H2/H4 beam configuration. Here we focus on describing the main items relevant for our study. Before the target a set of magnets define the incoming angle of the protons interacting with T2, which can vary from 0 up to 10~mrad (on average) depending on the desired H2/H4 configuration. The remaining protons ($\sim 30\%$) arising from the target are deviated towards a large collimating structure. It has 3.2~m of Iron and acts as a dump (designated ``TAX''), as shown in the lower panel of Fig.~\ref{fig:setup}. Subsequently, and given the slope of the secondary line especially in the vertical plane (see the top panel of Fig.~\ref{fig:setup}), the remaining particles produced at the dump are essentially absorbed by the various magnetic elements present in the environment and soil of length $\sim 500$ m, which act as shielding. In our calculations, we take a 0~mrad angle for the proton collisions on the target, and a 7~mrad angle for the protons that collide on the dump. Therefore, the flux of particles has two main components, as illustrated on the bottom panel in Fig.~\ref{fig:setup}.

As shown in Fig.~\ref{fig:setup}, the produced flux of BSM particles would reach both ProtoDUNE modules. In this work we will focus on the first detector (NP02), which is closer; however, a similar approach can be applied for the one downstream, and we have explicitly checked that the results obtained are very similar. Specifically, NP02 has a fiducial volume of $V_{det} = 6~\mathrm{m} \times 7~\mathrm{m} \times 6~\mathrm{m}$ and is located at a distance of $L=678$~m from the T2 target, as shown in Fig.~\ref{fig:setup}. As already mentioned, an important peculiarity of this experimental configuration is that there is no decay pipe available. In addition, the strong magnetic field that deviates the remaining protons after they hit the target will also deflect other charged particles (such as pions and kaons) with a much larger opening angle. Therefore the expected background from SM neutrinos at the detector will be considerably suppressed and will be neglected here. Instead, being on surface a significant background is expected from cosmic rays. However these may be reduced applying timing and kinematic cuts as discussed in more detail below.

\section{Results}
\label{sec:results}

As the SPS beam hits the T2 target, and subsequently the remaining protons hit the dump, the proton collisions create a plethora of unstable mesons, which may produce BSM particles as they decay. As outlined in the previous section, here we will focus on the production of new particles from the decays of short-lived mesons (namely, $\pi^0, \eta, \eta', \rho, \omega, J/\psi, D, D_s, \Upsilon$ and $B$ decays) as well as from $\tau$ decays. We extract the event distributions of the parent mesons from Pythia (v8.3.07)~\cite{Bierlich:2022pfr} using the \texttt{SoftQCD} flag. \pc{For $D, D_s$ mesons the \texttt{HardQCD} flag is used instead, as we found that this yields results that are more closely aligned with the simulations performed by the SHiP collaboration in Ref.~\cite{ship-note}, which were tuned to reproduce data on charm production from the E791 experiment~\cite{E791:1999oce}.} Note that this is generally conservative as we are not taking into account the production from secondary interactions in the target, which can lead to a non-negligible enhancement of the meson production rate (see for example Fig.~6 in Ref.~\cite{ship-note} for $D$ mesons). Thus, we expect our results to improve with a dedicated flux simulation that takes into consideration the full geometry of the experimental setup, which we leave for future work. Table~\ref{tab:mesons} summarizes the production rates for the parent particles considered in this work, normalized per PoT, for a beam with incident momentum of 400~GeV. 
\begin{table}[tb!]
    \centering
    \renewcommand{\arraystretch}{1.5}
    \begin{tabular}{|c|c|c|c|c|c|}
    \hline
       $\pi^0$  & $\eta$ & $\eta'$ & 
       $D$& $D_s$ & 
       $\tau$ \\
       $4.03$  & $0.46$ & $0.05$ & 
       $4.8\cdot10^{-4}$ & $1.4\cdot10^{-4}$ & 
       $7.4\cdot 10^{-6}$ 
        \\ \hline \hline 
       $\rho$& $\omega$ & $\phi$ & $J/\psi$ & 
       $B$ & $\Upsilon$ \\
       $0.54$ & $0.53$ & 0.019 & $4.4\cdot10^{-5}$ & 
       $1.2\cdot10^{-7}$ & $2.3\cdot10^{-8}$ \\
       \hline
    \hline
    \end{tabular}
    \caption{Production yield (normalized per PoT) for each of the parent particles considered in this work, see text for details. Note that the number of $\tau$ leptons receives contributions from direct production as well as from indirect production through $D_s \to \tau \nu$, the latter being dominant. }
    \label{tab:mesons}
\end{table}

Starting from the parent particle distributions, we then simulate their decays assuming either a two-body or a three-body process, depending on the parent particle and the scenario being considered. We then follow each BSM particle along its trajectory, keeping only those which intersect the detector fiducial volume (the flux \emph{accepted} by the detector). The expected number of events is computed from the accepted flux either taking into account the probability of decay inside the detector volume (in the case of unstable particles, which is related to their lifetime) or their probability to interact (in the case of stable particles, which is related to their interaction cross section). The relevant backgrounds will also be different, owing to the different signals expected in each case. Thus, in the remainder of this section, these two scenarios will be discussed separately. 

\subsection{Long-lived particles}
\label{sec:LLP}

For a long-lived particle $\Psi$ with mass $m_\Psi$ and lifetime $\tau_\Psi$ that is produced from a given parent meson\footnote{Although in this section we refer to mesons throughout the text, the expressions can be trivially generalized to the case of $\tau$ decays.} in the decay $M \to \Psi + \ldots$, the expected number of decays inside the detector can be computed as: 
\begin{widetext}
\begin{equation}
\label{eq:Ndec}
    N_{dec}^M = N_\mathrm{PoT} \, Y_{M} \, \mathrm{BR}(M \to \Psi) \, \int dS  \int dE_\Psi \,\mathcal{P}(c\tau_\Psi/m_\Psi, E_\Psi, \Omega_\Psi) \, \dfrac{dn^{M\to \Psi}}{dE_\Psi dS} \, ,
\end{equation}
\end{widetext}
where $\mathrm{BR}(M \to \Psi)$ is the production branching ratio in the decay of meson $M$, $N_\mathrm{PoT}$ is the number of protons on target integrated over a given data taking period, $Y_{M}$ is the meson production yield (provided in Tab.~\ref{tab:mesons}), and $\frac{dn^{M\to \Psi}}{dE_\Psi dS}$ stands for the number of particles produced from the decay $M \to \Psi$ with energy $E_\Psi$ entering the detector through a differential surface $dS$, with a trajectory defined by the solid angle $\Omega_\Psi$. The decay probability inside the detector volume reads:
\begin{equation}
    \mathcal{P} = e^\frac{-\ell_{det} }{L_\Psi} \left( 1 - e^\frac{-\Delta \ell_{det}}{L_\Psi}\right) \, ,
    \label{eq:Prob}
\end{equation}
where $\ell_{det}$ is the length of the trajectory before the particle enters the detector, and $\Delta \ell_{det}$ is the length of the trajectory inside the detector (note that both quantities depend on the solid angle $\Omega_\Psi$). On the other hand, $L_\Psi$ is the boosted decay length in the laboratory frame: $ L_\Psi = \gamma_\Psi \beta_\Psi c\tau_\Psi \simeq c\tau_\Psi E_\Psi/m_\Psi $.

In the limit of small couplings, the lifetime of the particle will be much longer than both the distance to the detector and the length traveled by the particle inside it. In this limit it is illustrative to consider the case where the dependence of $\ell_{det}$ and $\Delta \ell_{det}$ with the solid angle is neglected, which leads to:
\begin{equation}
    N_{dec}^M \simeq N_\mathrm{PoT} \, Y_{M} \, \mathrm{BR} (M \to \Psi) V_{det} \int
    \,\frac{dE_\Psi}{L_\Psi}\, \left\langle \dfrac{dn^{M\to \Psi}}{dE_\Psi dS} \right\rangle . 
\nonumber 
\end{equation}
Here, our notation $\langle \ldots \rangle\equiv \frac{1}{S} \int \ldots dS $ indicates the average taken within the detector size, for a given energy. As can be seen, the number of events approximately scales with the volume of the detector, $V_{det} = S_{det} \Delta \ell_{det}$. Thus, the number of decays will be enhanced within the ProtoDUNE detectors, thanks to their large fiducial volume. Although this expression is useful to understand the behaviour of the results, we stress again that our numerical calculation of the accepted flux does take into account the detector location, shape, and angle with respect to the beam direction as outlined in Sec.~\ref{sec:setup}. 

The final number of events needs to take into account the branching ratio of the decay into a visible final state and therefore, the results of Eq.~\ref{eq:Ndec} should be multiplied by the corresponding branching ratio, $\mathrm{BR}(\Psi \to \mathrm{Visible})$, times the detection efficiency for a given final state, $\epsilon_{det}$. Therefore, for a given mass, the observable number of events will depend on three model-dependent quantities: the branching ratio for the production of $\Psi$, its branching ratio into visible states, and its lifetime in the rest frame. While in specific models these three quantities may be related, it is useful to derive model-independent sensitivities that can be easily recasted to specific scenarios \pc{(see e.g. Refs.~\cite{Arguelles:2019ziu,Coloma:2019htx,Coloma:2022hlv} for recent examples).}

\begin{center}
    \begin{figure}[ht!]
        \centering
        \includegraphics[width=0.95\columnwidth]{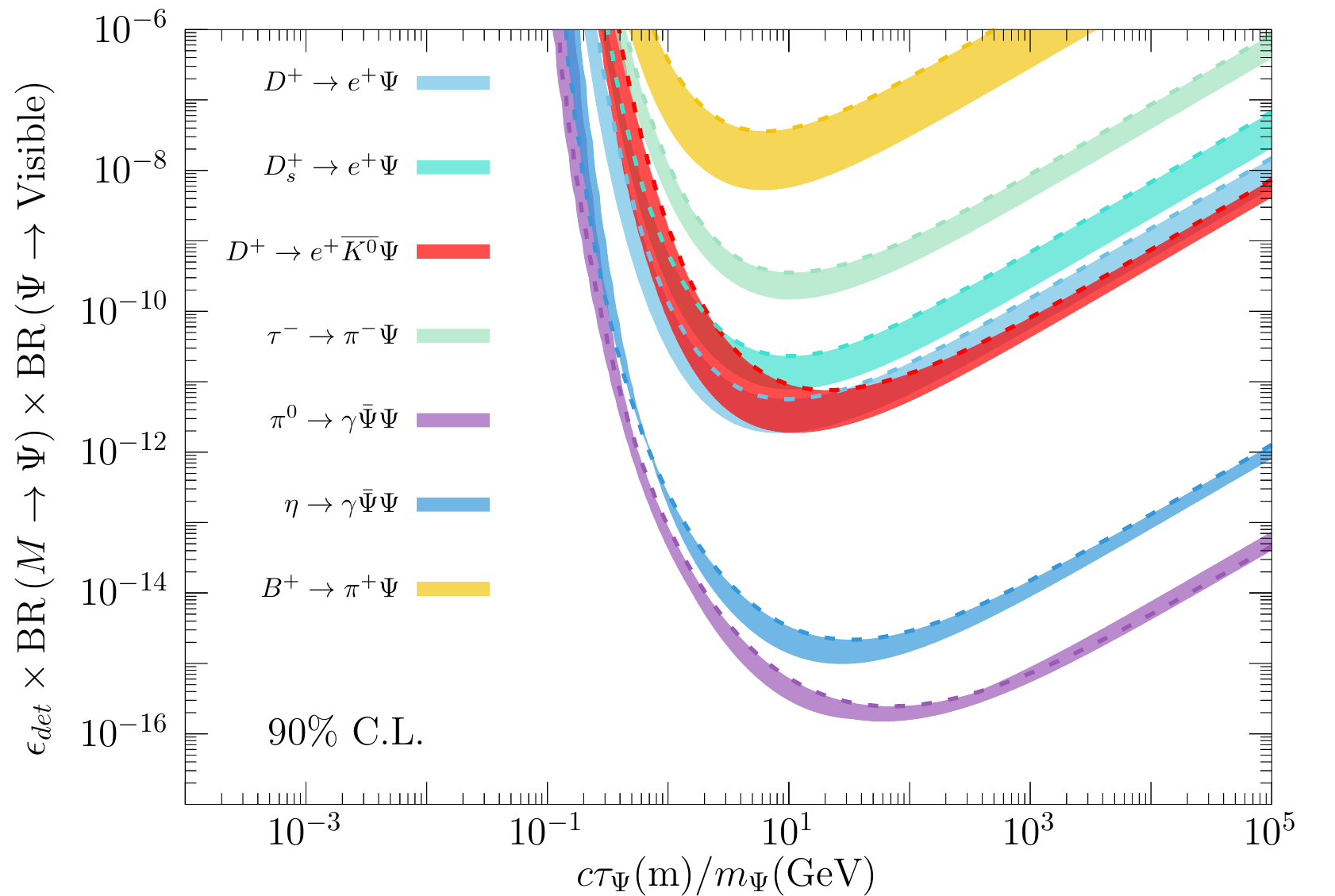}
        \caption{Expected sensitivity to long-lived particles in the model-independent scenario, assuming the branching ratios and lifetime at rest of the long-lived particle are uncorrelated. The region above each dashed line would lead to a number of events above 2.44 in 5 years of data taking, which in the absence of backgrounds would correspond to 90\% confidence level (C.L.). The width of the bands indicate the variation in the results for masses of the long-lived particle between 10~MeV and up to the production threshold in each case, see text for details.}
        \label{fig:BR-ctau}
    \end{figure}
\end{center}

Our results are shown in Fig.~\ref{fig:BR-ctau}, where we have computed the regions where the number of signal events would exceed 2.44 in 5~years of data taking, for different fermion production mechanisms as a function of $m_\Psi$. This would correspond to a 90\% confidence level (C.L.) sensitivity in the absence of backgrounds~\cite{Feldman:1997qc}. Regarding the expected ProtoDUNE efficiencies, the reconstruction efficiency for energetic particles is expected to be above 80\% (see Sec. 4.3 in Ref.~\cite{DUNE:2020ypp}). Nevertheless, the final efficiency achievable will depend on the cuts applied to reduce the backgrounds, which cannot be estimated without a dedicated analysis. \pc{
In particular, we note that the energies of the LLP are expected to be in the range of tens of GeV and consequently their decay products will also be very energetic. At such high energies, the ProtoDUNE modules are well above threshold and no large changes in efficiency are therefore expected. Therefore, we leave our results in this case as a function of the efficiency, which we assume to be constant above threshold.} The lines are shown as a function of $c\tau_\Psi / m_\Psi$, as this ratio determines the point that maximizes the decay probability within the detector in Eq.~\ref{eq:Prob}, which at first approximation leads to an optimal sensitivity. However, since the boost kinematics will be slightly different depending on the values of $m_\Psi$, we obtain a different detector acceptance for different masses (see e.g., the discussion in Ref.~\cite{Coloma:2020lgy}) which leads to slight variations in the results. This is indicated by the width of each band, which has been obtained scanning masses between 10~MeV (indicated by the dashed lines) and the largest mass kinematically accessible for each production channel.

\begin{center}
    \begin{figure*}[ht!]
        \centering
        \includegraphics[width=0.95\textwidth]{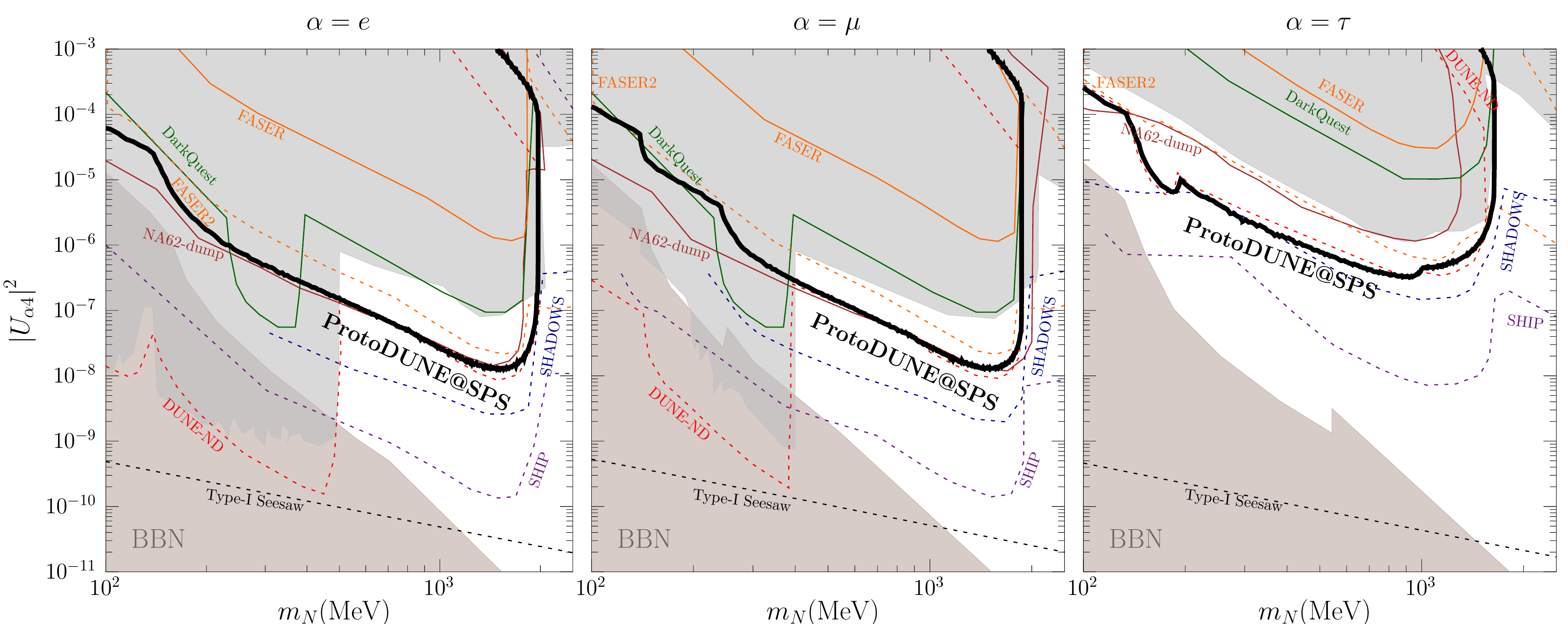}
        \caption{Expected sensitivity to Heavy Neutral Leptons (HNLs) at $90\%$ C.L., as a function of the HNL mass. In each panel, results are obtained setting the remaining mixing parameters to zero, and assuming backgrounds can be reduced to a negligible level. Our results are given by the solid black lines, while current constraints are indicated by the shaded gray areas (extracted from~\cite{Agrawal:2021dbo} as well as the phenomenological recasts of results from CHARM~\cite{Boiarska:2021yho} and BEBC~\cite{Barouki:2022bkt}). The sensitivities of other beam-dump experiments are also shown for comparison, where solid (dashed) lines correspond to experiments taking data on a similar (longer) timescale,
        see text for details. The dashed black line corresponds to the naive seesaw scaling and should be considered only as indicative. It corresponds to $|U_{\alpha 4}|^2 =\sqrt{\Delta m^2_\mathrm{atm}}/m_N$, where $\Delta m^2_\mathrm{atm}$ is the light neutrino atmospheric mass-squared difference. The brown shaded area is disfavored by BBN bounds extracted from~\cite{Abdullahi:2022jlv,Boyarsky:2020dzc}. \label{fig:HNL}}
    \end{figure*}
\end{center}

Let us now focus on a specific model which is well-motivated from the theoretical point of view, such as the Heavy Neutral Lepton (HNL) scenario. In particular, HNLs arise in low-scale Seesaw models, which can generate neutrino masses~\cite{Branco:1988ex,Kersten:2007vk,Abada:2007ux,Gavela:2009cd} and the baryon asymmetry of the Universe~\cite{Akhmedov:1998qx,Asaka:2005pn}, thus solving two of the main open problems in the SM. At the same time, HNLs with masses in the GeV range are relatively hard to produce at fixed-target experiments since their main production mechanism is through $D, D_s$ and $\tau$ decays, which are hard to produce in the laboratory. Therefore, this mass region is far less constrained that HNLs with masses below the kaon mass (for a recent review see Ref.~\cite{Abdullahi:2022jlv}). In turn, we expect our setup to yield competitive constraints thanks to the high beam energy available at the SPS. In order to compare to current experimental constraints and future sensitivities, we will consider a simplified scenario with one \jb{Dirac}\footnote{\jb{The main difference with respect to the Majorana case is a factor two in the decay rate (see e.g. the related discussion in Ref.~\cite{Ballett:2019bgd}), which would lead to a very minor difference on the sensitivity. }} HNL of mass $m_N$ that mixes exclusively with one SM neutrino of a given flavor.
The relevant portion of the Lagrangian reads:
\begin{equation}
    \mathcal{L} \supset - \frac{m_W}{v} \overline N U_{\alpha 4}^*\gamma^\mu l_{L \alpha} W^+_\mu
- \frac{m_Z}{\sqrt 2 v} \overline N U_{\alpha 4}^* \gamma^\mu \nu_{L \alpha} Z_\mu \, ,
\end{equation}
where $l_\alpha$ and $\nu_\alpha$ stand for the charged lepton and light neutrino of flavor $\alpha \equiv e,\mu,\tau$, while $v$ stands for the Higgs vacuum expectation value, $m_Z$ ($m_W$) is the mass of the $Z$ ($W$) boson, and $U_{\alpha 4}$ indicates the mixing matrix elements between the HNL and the light neutrinos. 
In this scenario the HNL production branching ratio and its decay width will be strongly correlated and depend on the HNL mass and its mixing with the light neutrinos. In our calculations, we take the HNL production branching ratios and decay widths from Ref.~\cite{Coloma:2020lgy} (see also Ref.~\cite{Bondarenko:2018ptm}). \jb{Just as a reference, for a HNL with mass $m_N \sim 1$~GeV and mixing with the active neutrinos $|U_{\alpha 4}|^2 \sim 10^{-4}$, the lifetime is approximately $c\tau_N\approx \mathcal{O}(10-20)\,\rm{m}$, depending on the neutrino flavor that mixes with the HNL. As the produced HNL flux is quite energetic (with typical energies in the ballpark of $\sim 50~\mathrm{GeV}$), this leads to boosted decay lengths in the range between 500~m and 1~km.}
Once produced, the HNL will decay back to SM particles (mesons and leptons) through its mixing. In the following, we will consider its decays to the following final states: $N \rightarrow \nu e e, \nu \mu \mu, \nu e \mu, e \pi, \mu \pi$ and $\nu \pi^0$, which would be easily identifiable in the ProtoDUNE detectors thanks to their excellent particle identification (ID) capabilities. As for the backgrounds, since the neutrino production in the beam is heavily reduced (thanks to the dump, as explained in the previous section) we expect the largest contribution to come from cosmic rays. However, we think that these can also be reduced to a negligible level due to several factors. First, angular and timing cuts could be applied to remove those events that are not coming from the direction of the target within a beam spill time window. Second, in the case of fully visible final states (such as $N \rightarrow \mu \pi$) the event can be fully reconstructed~\cite{MicroBooNE:2017xvs, DUNE:2020cqd}, which would offer additional handles to suppress the background with respect to the signal (an example of a similar search at MicroBooNE can be found in Ref.~\cite{MicroBooNE:2019izn}, see also Refs.~\cite{Ballett:2019bgd,Breitbach:2021gvv} for sensitivity studies using signal-to-background discrimination techniques at the DUNE near detectors). Therefore our results for this scenario have been obtained assuming a negligible background level, leaving a detailed calculation of the expected background for future work. \pc{In this case we have assumed a perfect detector efficiency, $\epsilon_{det}=1$, but our results can be easily recasted for a different value.} The expected sensitivity, shown in Fig.~\ref{fig:HNL} for an inclusive search using the decay channels indicated above, indicates that this setup would significantly improve over current constraints in the mass window above 400~MeV, and would be competitive (or even better) than other facilities planned on a similar timescale at CERN, such as NA62-dump~\cite{Collaboration:2691873,Beacham:2019nyx}, FASER~\cite{FASER:2018eoc} or DarkQuest~\cite{Batell:2020vqn}, indicated by solid lines. Most notably, the sensitivity of our setup lies approximately in the same ballpark as proposals on a longer timescale such as, e.g., FASER2~\cite{FASER:2018eoc} 
or the DUNE ND-GAr~\cite{Coloma:2020lgy} 
(which belongs to Phase II of the DUNE experiment).
For comparison, in Fig.~\ref{fig:HNL} we also show the future sensitivity of SHIP~\cite{Aberle:2839677} and SHADOWS~\cite{Alviggi:2839484}. The corresponding line for HIKE~\cite{HIKE:2022qra} (within the mass range shown here) is very similar to that of SHADOWS and therefore is not shown.
An overview of the estimated timeline for the experiments listed above can be found e.g. in Ref.~\cite{timelineExps,Antel:2023hkf}. 

Finally, note that the results shown in Fig.~\ref{fig:HNL} are obtained for an inclusive search (including a number of final states, as listed above), in order to ease comparison with the literature. However, experimental searches targeting different HNL decay modes may be subject to different optimization strategies and background discrimination techniques. This may affect the final efficiencies and sensitivities differently for each decay channel. Therefore, in App.~\ref{app} we also provide the expected sensitivities separately for each decay mode of the HNL.

\subsection{Stable particles}
\label{sec:stable}

\begin{center}
    \begin{figure*}[t!]
        \centering
        \includegraphics[width=0.32\textwidth]{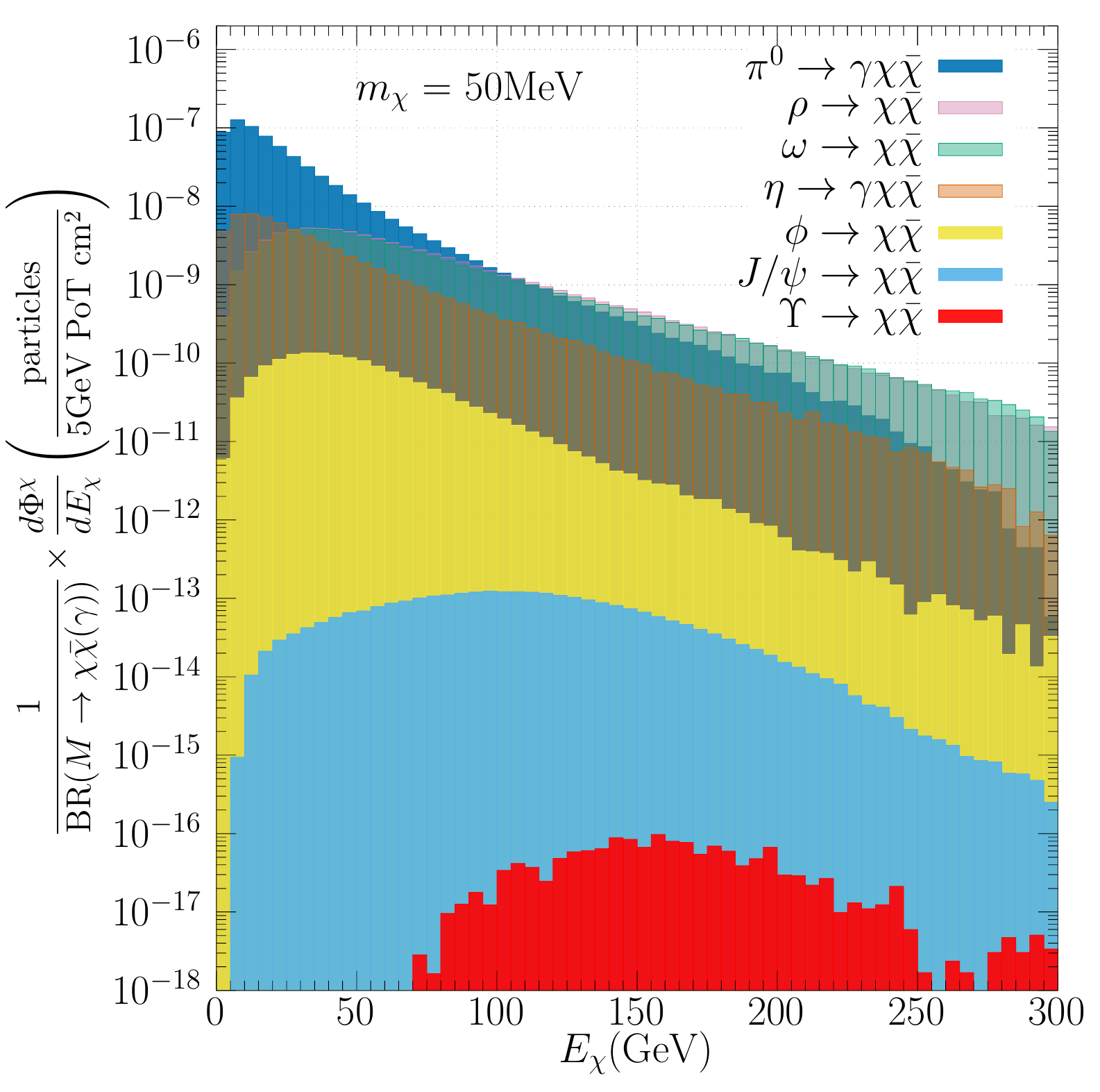}
        \includegraphics[width=0.32\textwidth]{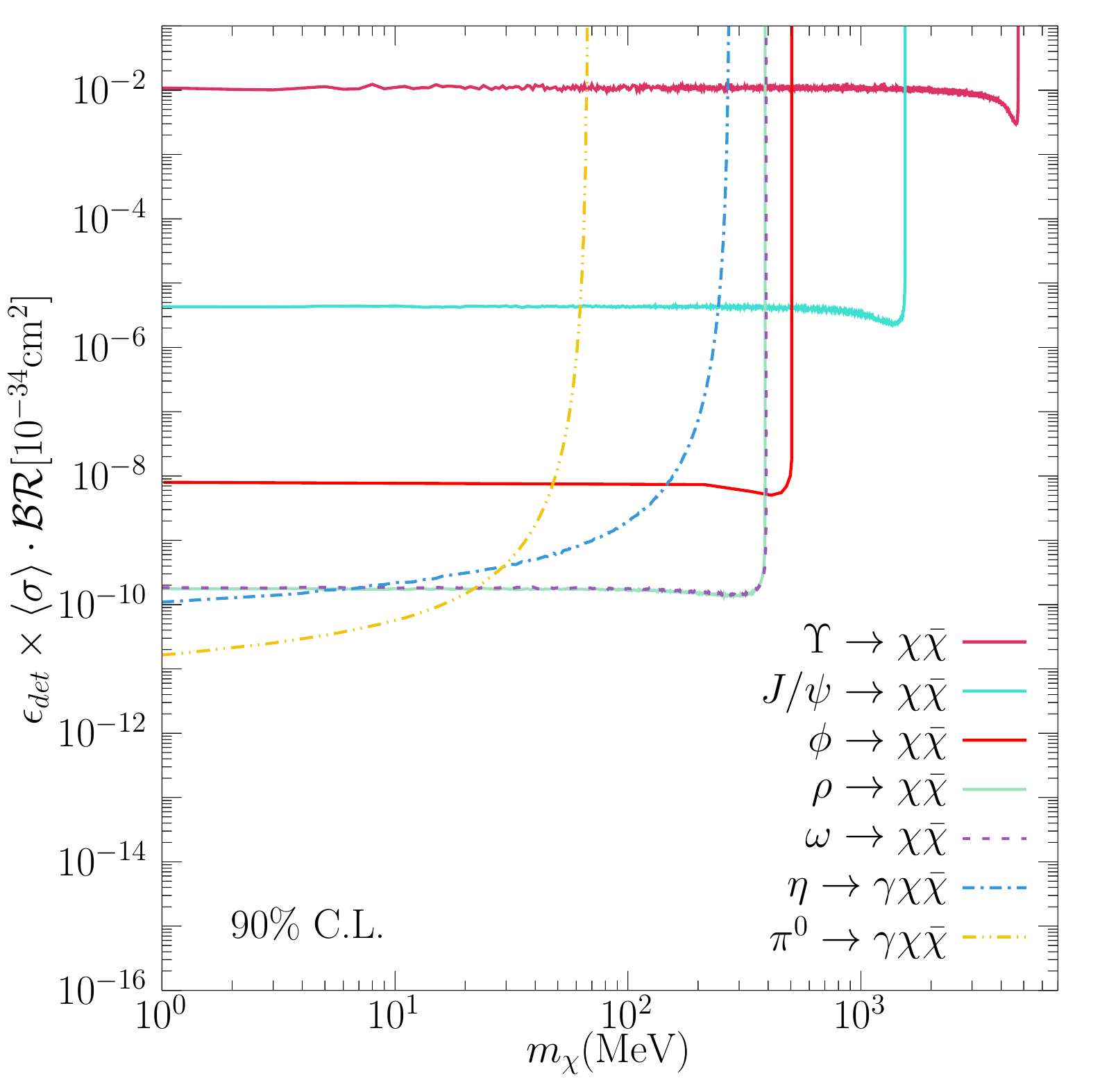}
        \includegraphics[width=0.32\textwidth]{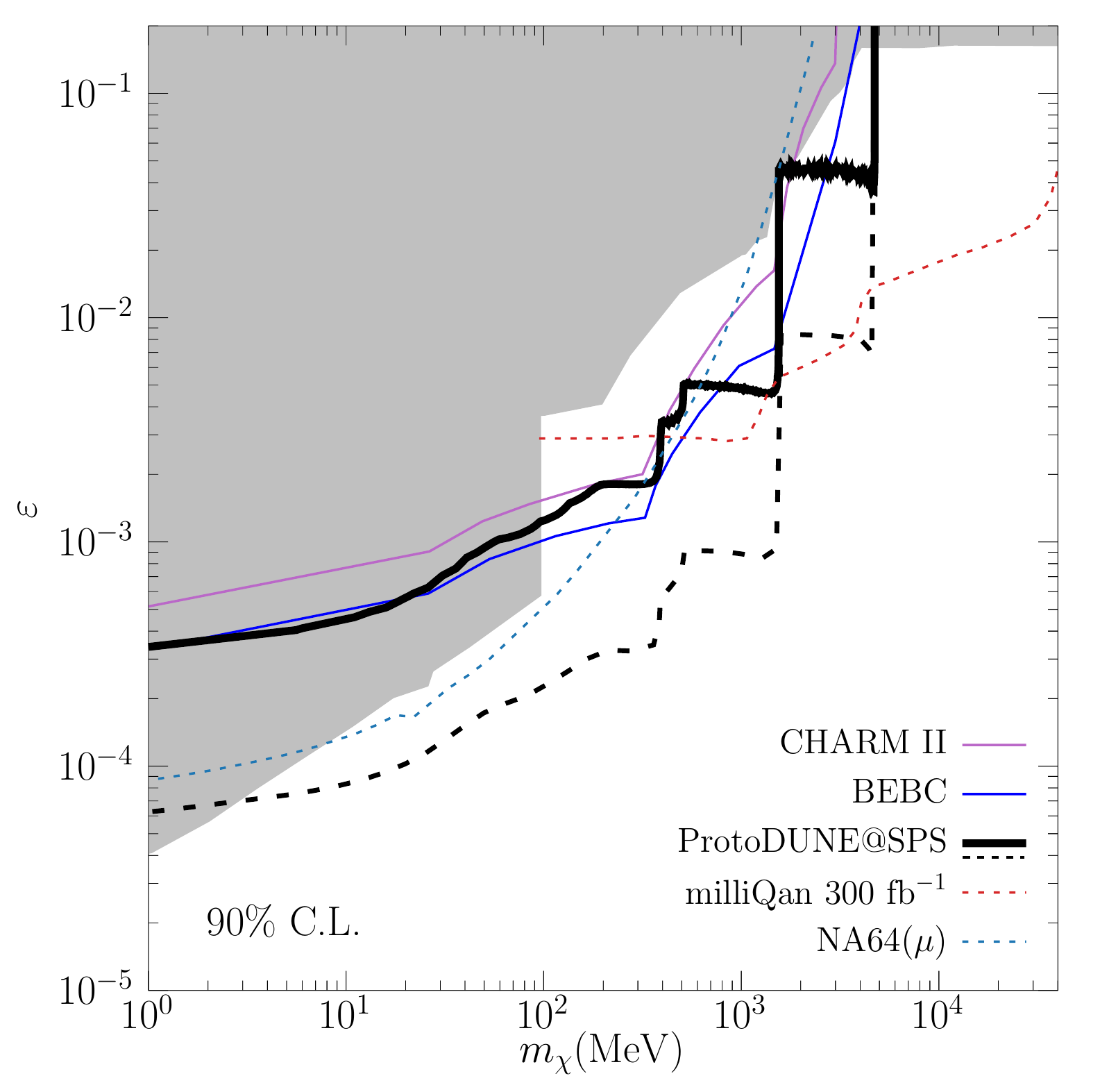}
        \caption{Expected fluxes and sensitivity to stable, weakly interacting particles. \textit{Left panel}: Flux of stable particles that would enter the fiducial volume of the ProtoDUNE detectors as a function of energy, for different production mechanisms and parent particles, as indicated in the legend. \textit{Middle panel}: Sensitivity expected to stable particles, for the model-independent case, assuming no backgrounds and perfect detection efficiency. Here, $\langle \sigma \rangle$ is the average cross section defined in Eq.~\ref{eq:cross_convolution}, while $m_\chi$ is the mass of the stable particle, \jb{and $\mathcal{BR}$ is the production branching ratio stripped of the phase space suppression factor, see main text for details.} \textit{Right panel}: Sensitivity expected for the millicharged particle scenario as a function of its mass $m_\chi$ (solid black line), see text for details. Since the setup is background-limited, the dashed black line indicates the ultimate sensitivity achievable if backgrounds could be significantly reduced, see text for details. We compare our results to previous constraints from SLAC~\cite{Prinz:1998ua}, LSND and MiniBooNE~\cite{Magill:2018tbb}, ArgoNeuT~\cite{ArgoNeuT:2019ckq}, miliQan~\cite{Ball:2020dnx} and LEP~\cite{Davidson:2000hf} (gray filled area); recent recasts of results from CHARM II and BEBC~\cite{Marocco:2020dqu} (solid violet and blue lines respectively); and the future expected sensitivity of NA$64\mu$~\cite{Gninenko:2640930,Gninenko:2018ter} (dashed light blue line) and milliQan with $300$ fb$^{-1}$ of integrated luminosity~\cite{Ball:2016zrp} (dashed red line), with a similar timescale to our proposal. \label{fig:MCP}}
    \end{figure*}
\end{center}

Being filled with LAr, the ProtoDUNE detectors would also be sensitive to the interactions of BSM particles. While unstable particles may also be searched for in this manner, here we focus on the case of stable particles for simplicity. 
Let us assume that the leading production mechanism for a light stable particle $\chi$ comes from a given parent meson $M \to \chi \bar\chi + \ldots$, which will interact as it arrives to the ProtoDUNE modules with an interaction cross section $\sigma$. An average interaction cross section may then be defined as:
\begin{equation}
\label{eq:cross_convolution}
\begin{aligned}
\langle\sigma\rangle= & \frac{1}{\Phi^\chi} \int_0^{\infty}\int_{T^{\min }}^{T^{\text {max }}} \frac{d\sigma}{dT}\left(E_\chi, \left\{X\right\}\right) \frac{d \Phi^\chi}{d E_\chi} d T  d E_\chi 
\end{aligned}
\end{equation}
where $T$ is the recoil energy of the electron, while $E_\chi$ is the energy of the particle, $\left\{X\right\}$ are the model parameters, and $\Phi^{\chi}$ is the flux of the incoming $\chi$ particles which trajectories intersect the detector (in units of $\text{PoT}^{-1}\text{cm}^{-2}$, and averaged over all possible trajectories). Note that the cross section here is integrated between the minimum observable recoil energy (which depends on the detector technology) and the maximum achievable recoil (which depends on kinematics). The number of events can then be written in terms of the average interaction cross section, as:
\begin{equation}
\label{eq:Nevsigma}
    N_{ev} = \epsilon_{det} \, N_{trg} \, \langle \sigma \rangle \, \Phi^{\chi} \, N_{\text{PoT}} \, ,
\end{equation}
where $N_{trg}$ is the number of targets relevant for the interaction (e.g., electrons, or nuclei) contained in the fiducial volume of the detector. Notice that the flux $\Phi^\chi$ depends on the production branching ratio, which is also a function of the model parameters \pc{and includes a phase space suppression which is a function of the mass of the parent and the daughter particles, $\mathrm{PS} \equiv \mathrm{PS}(m_\chi, m_M)$, defined in such a way that in the limit $m_\chi/m_M \to 0$, $\mathrm{PS} \to 1$. Equation~\eqref{eq:Nevsigma} can be rewritten to include this dependence explicitly, as:
\begin{widetext}
\begin{equation}
\label{eq:Nevsigma-2}
    N_{ev} = \epsilon_{det} \, N_{trg} \, \left[\langle \sigma \rangle \cdot \mathcal{BR}\right] \; \mathrm{PS}\left(m_\chi, m_M\right) \, \frac{\Phi^{\chi}}{\mathrm{BR}(M \to \chi \bar\chi \ldots)} \, N_{\text{PoT}} \, ,
\end{equation}
\end{widetext}
where $\mathrm{BR}(M \to \chi \bar\chi \ldots) = \mathcal{BR}\cdot\mathrm{PS}$, defining $\mathcal{BR}$ as the branching ratio stripped of the phase space suppression factor. Using this formalism, the dependence on the model parameters is contained in the quantity $\langle\sigma \rangle \cdot \mathcal{BR}$
, for which model-independent sensitivity limits can be obtained, as shown in the middle panel in Fig.~\ref{fig:MCP}.} For simplicity, in this figure we have assumed no \pc{backgrounds; we also leave our results as a function of the detection efficiency (which we again assume to be constant above threshold).} In addition, these results can be easily recasted to a specific model using the fluxes provided in the left panel of Fig.~\ref{fig:MCP}. In particular, we note that our sensitivity regions shown in the middle panel of Fig.~\ref{fig:MCP} would also be applicable to any BSM scenario inducing the production mechanisms listed above, including (but not restricted to) millicharged  particles (MCPs) or a dark portal through a massive vector mediator~\cite{Okun:1982xi,Holdom:1985ag}.

In order to compare to current bounds from other experiments, let us now focus on a particular scenario. Specifically we consider the case of MCPs, which arise in certain BSM scenarios from the mixing between the SM photon and a massless dark photon~\cite{Holdom:1985ag}. MCPs are fermions with an effective electric charge $\varepsilon e$ ($e$ being the electric charge of the electron) and mass $m_\chi$. As they arrive to the detector, these would lead to an excess of electron recoils. The differential electron scattering cross section for this process is given by~\cite{Harnik:2019zee}:
\begin{equation}\label{eq:crossMCP}
    \frac{d \sigma}{d T}=\pi \alpha^2 \varepsilon^2 \frac{2 E_\chi^2 m_e+T^2 m_e-T\left(m_\chi^2+m_e\left(2 E_\chi+m_e\right)\right)}{T^2\left(E_\chi^2-m_\chi^2\right) m_e^2}
\end{equation}
%
where $m_e$ is the electron mass. As can be seen, in the limit $E_\chi \gg T, m_e, m_\chi$ it is enhanced at low recoil energies. Therefore, we naively expect the ProtoDUNE detector to be highly sensitive to such a signal, thanks to the low thresholds achievable at LAr TPCs. While the detection threshold for electron recoils at \pc{ProtoDUNE is expected to be around 10-30~MeV~\cite{Chatterjee:2018mej,DUNE:2020ypp, DUNE:2022meu}, we assume 30~MeV in our calculations~\cite{DUNE:2022meu,MicroBooNE:2018dfn}.} Thus, in the limit of small MCP masses, and taking $E_\chi \gg T, m_e, m_\chi$, the size of the interaction cross section for this model can be estimated as
\begin{equation}
   \sigma \sim \varepsilon^2 \, \left( \frac{30~\mathrm{MeV}}{T_\mathrm{min}}\right) \, 10^{-26}  ~\mathrm{cm^{-2}} , \nonumber
\end{equation}
which implies
\begin{align}
   \frac{\langle \sigma \rangle\times \mathrm{BR} }{10^{-26}~\mathrm{cm^2}} & \sim \mathrm{BR}(\pi^0 \to \gamma \chi \bar\chi) \, \varepsilon^2 \, \left( \frac{30~\mathrm{MeV}}{T_\mathrm{min}}\right) \,  \nonumber \\  
   & \sim \mathrm{BR}(\pi^0 \to \gamma e^- e^+) \, \varepsilon^4 \, \left( \frac{30~\mathrm{MeV}}{T_\mathrm{min}}\right) \, . \label{eq:estimate} 
\end{align}
According to the middle panel in Fig.~\ref{fig:MCP}, and taking $\mathrm{BR}(\pi^0 \to \gamma e^+ e^-) = 1.174\%$~\cite{Workman:2022ynf}, this implies that our setup could potentially be sensitive to values of $\varepsilon$ as low as $\sim 5\times 10^{-5}$, in the absence of backgrounds. 

However, for electron recoils at such low energies we expect a significant background from cosmic ray interactions in the detector. Unlike in the case of long-lived particles, these will be harder to disentangle from the signal events and therefore we expect these to significantly reduce our final sensitivity to this scenario. 
Thus, the main  background will be cosmogenic muons that are energetic enough ($E_{\mu}\gtrsim \text{400~GeV}$) to penetrate the fiducial volume but do not leave a distinguishable muon-like track. Following Ref.~\cite{Chatterjee:2018mej}, here we define a slightly smaller fiducial volume of $6~\mathrm{m}\times7~\mathrm{m}\times5.65~\mathrm{m}$. Within this volume, the total number of muons with energies above 400 GeV is approximately $4 \times 10^{11}$ per year~\cite{Chatterjee:2018mej}. However, only $30\%$ of these muons occur simultaneously with the spill (see Sec.~\ref{sec:setup}), and only around $0.1\%$ do not leave a muon-like track~\cite{Chatterjee:2018mej}. Furthermore, the incoming muon flux has an angular dependence that varies as $\sim \sin^2\theta$, where $\theta$ is the angle with respect to the horizontal. 
To further reduce this background, we apply an angular cut of $10^\circ$ above the horizontal. 
After applying these conservative cuts, we are left with approximately $2\cdot10^6$ background events per year. 

Detailed measurements of the cosmic ray background can be performed using beam-off data at ProtoDUNE and, therefore, it is reasonable to assume that our signal significance will be mainly limited by the statistical error on the background, while systematic uncertainties will be subleading. Thus, the sensitivity limit for this scenario can be computed using a Gaussian $\chi^2$, as:
\[
\chi^2 =\left( \frac{N_{ev} - N_{bg}}{\sqrt{N_{bg}}}\right)^2 \, ,
\]
which is computed using just the total number of events. We again stress here that our results are conservative as binning in recoil energy may offer additional handles to enhance the signal significance, provided that the background shows a different dependence with recoil energy than the signal. \pc{The expected sensitivity of our setup is shown in the right panel of Fig.~\ref{fig:MCP}, assuming perfect detection efficiency. We compare our results to previous limits in the literature, and to the} expected sensitivity of experiments with a similar timescale such as NA$64\mu$~\cite{Gninenko:2640930} and milliQan (with $300$ fb$^{-1}$ of integrated luminosity, taken from Ref.~\cite{Ball:2016zrp}). Note that other future experiments such as JUNO, SHiP or DUNE, will also deliver very interesting limits for this scenario~\cite{Magill:2018tbb, ArguellesDelgado:2021lek}. As can be seen, using just the total event rates and even with our conservative estimates regarding luminosity and the treatment of backgrounds, we expect the setup to be competitive with the most stringent limits in the parameter space for a wide range of masses. For comparison, we also show the potential improvement if a reduction of the background was possible significantly below the value considered here. This is indicated by the dashed black lines, obtained considering perfect background rejection. Although this is not realistic, it serves as an indication of the room for improvement for this setup, which for this scenario is background-limited. We note that our numerical calculations show perfect agreement with our naive estimate for the sensitivity based on Eq.~\ref{eq:estimate}.

\section{Summary and conclusions}
\label{sec:summary}

Given their location at CERN, the ProtoDUNE detectors may be exposed to a flux of new particles generated after the collision of 400~GeV protons, extracted from the SPS accelerator, with the T2 target (see Fig.~\ref{fig:setup}). In this Letter, we have explored the possibility of using such a setup to search for BSM weakly interacting particles in a beam-dump configuration. We have shown that it offers the opportunity to search for both long-lived unstable particles and stable particles, thanks to the large fiducial volume of the ProtoDUNE modules and to the high density of liquid Argon, respectively. Additional advantages of this setup include the absence of a decay pipe, which leads to a strong suppression of the beam-coincident neutrino events, and the excellent particle ID and tracking capabilities of LAr TPCs, required to suppress the cosmic-ray induced background. Our results show that the expected sensitivity goes considerably beyond current constraints for two representative examples (Heavy Neutral Leptons and millicharged particles) using facilities that are already in place at CERN, without interfering with the experimental program in the North Area, and within a relatively short timescale. However the possibilities offered by this setup are much wider, as it may also be used to search for additional weakly interacting particles such as dark photons, dark scalars, axion-like particles, or light dark matter. To illustrate its reach, we have also shown the expected sensitivity of the setup in a model-independent fashion that allows our results to be easily recasted to particular NP models involving either unstable or stable new states. Finally, we would like to remark that while our results have been derived under generally conservative assumptions, a dedicated analysis is required in order to determine the expected background levels and detector efficiencies achievable for such a setup. In particular, the study of a new trigger algorithm optimised for the beam-dump configuration is essential to fully determine the potential of this setup.

\vspace{0.7cm}
\textit{\textbf{Acknowledgements.}} We are very grateful to Nikolaos Charitonidis, Sylvain Girod and Vincent Clerc from the CERN BE-EA group for very useful discussions and insights on the SPS North Area layout and H2/H4 beamlines configurations. We warmly thank Paolo Crivelli and Sergei Gninenko for very useful discussions. We also warmly thank Francesco Pietropaolo for his useful feedback and reading of the manuscript, and Sara Bianco for pointing out an inconsistency in one of our plots in a previous version of the manuscript. This work has received partial support from the European Union's Horizon 2020 research and innovation programme under the Marie Sk\l odowska-Curie grant agreement No 860881-HIDDeN and the Marie Skłodowska-Curie Staff Exchange  grant agreement No 101086085 – ASYMMETRY. PC acknowledges partial financial support by the Spanish Research Agency (Agencia Estatal de Investigaci\'on) through the grant IFT Centro de Excelencia Severo Ochoa No CEX2020-001007-S and by the grant PID2019-108892RB-I00 funded by MCIN/AEI/ 10.13039/501100011033. She is also supported by Grant RYC2018-024240-I, funded by MCIN/AEI/ 10.13039/501100011033 and by ``ESF Investing in your future''. JLP and SU acknowledge support from Generalitat Valenciana through the plan GenT program (CIDEGENT/2018/019) and from the Spanish Ministerio de Ciencia e Innovacion through the project PID2020-113644GB-I00. JLP also acknowledges financial support by the Spanish Research Agency (Agencia Estatal de Investigaci\'on) through the project CNS2022-136013. The work of LMB is supported by SNSF Grant No. 186158 (Switzerland), RyC-030551-I, and PID2021-123955NA-100 funded by MCIN/AEI/ 10.13039/501100011033/FEDER, UE (Spain). The authors acknowledge use of the HPC facilities at the IFIC (SOM cluster) and at the IFT (Hydra Cluster).
\\

\appendix
\section{Non-inclusive sensitivities to HNLs}
\label{app}

\begin{figure*}[ht!]
\begin{center}
  \includegraphics[width=0.85\textwidth]{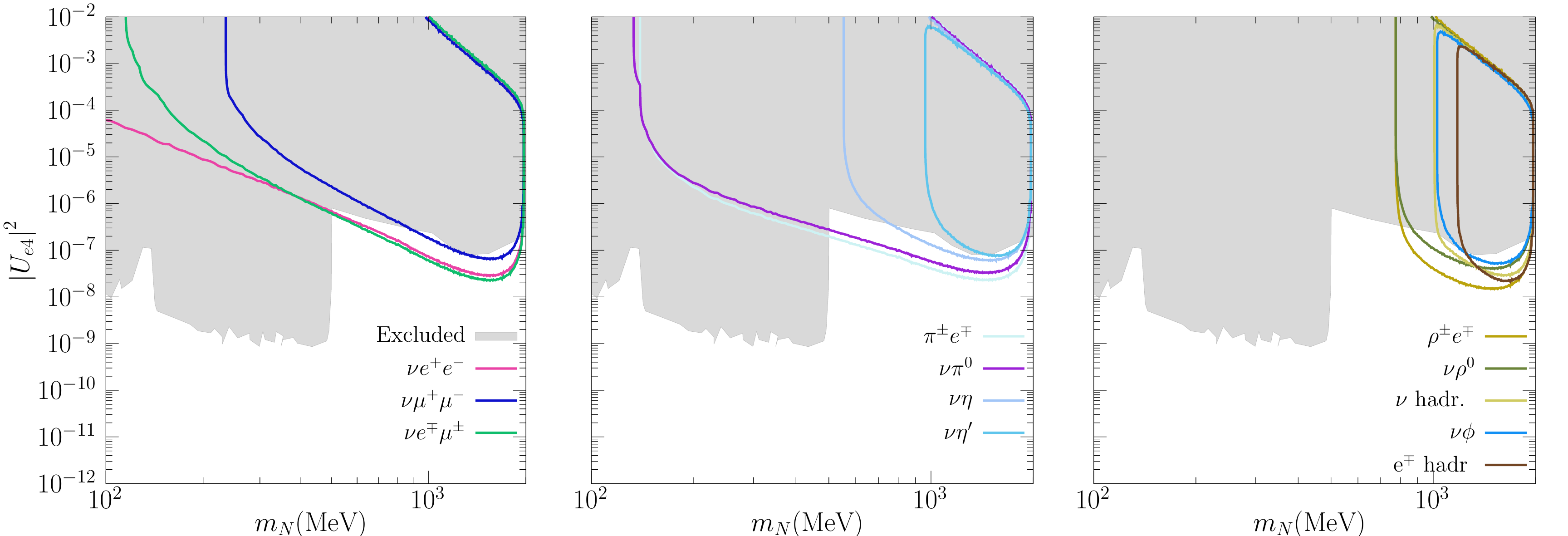}
  \includegraphics[width=0.85\textwidth]{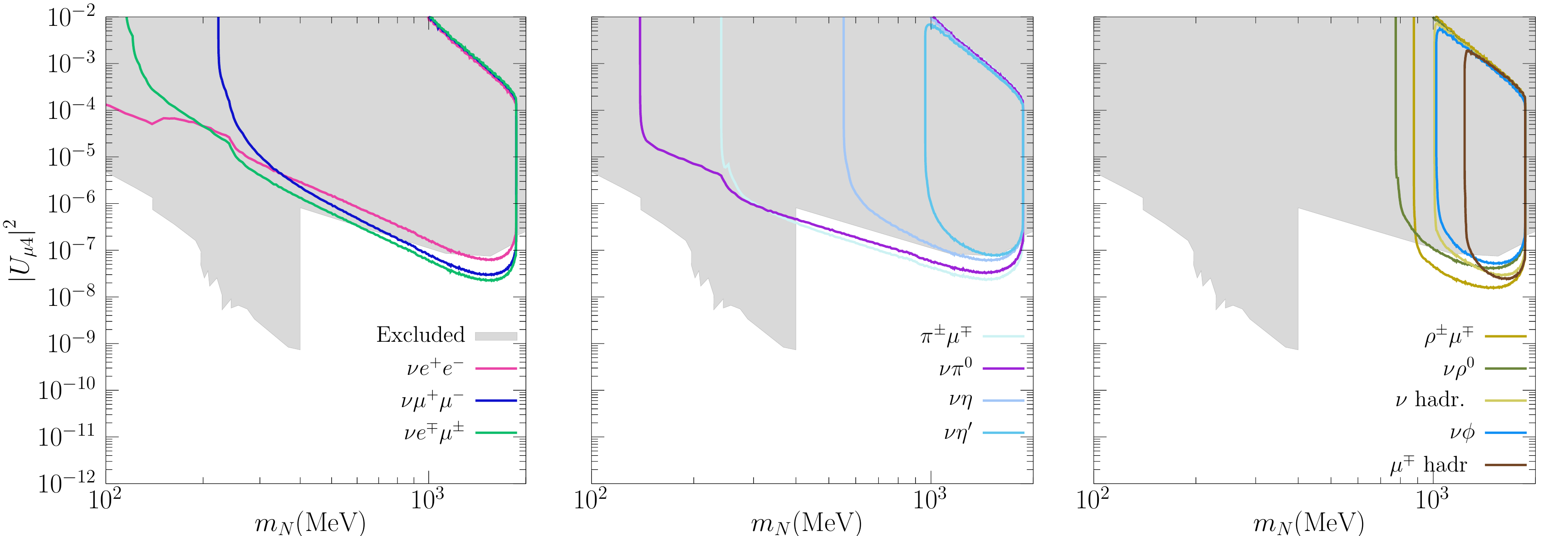}
  \includegraphics[width=0.85\textwidth]{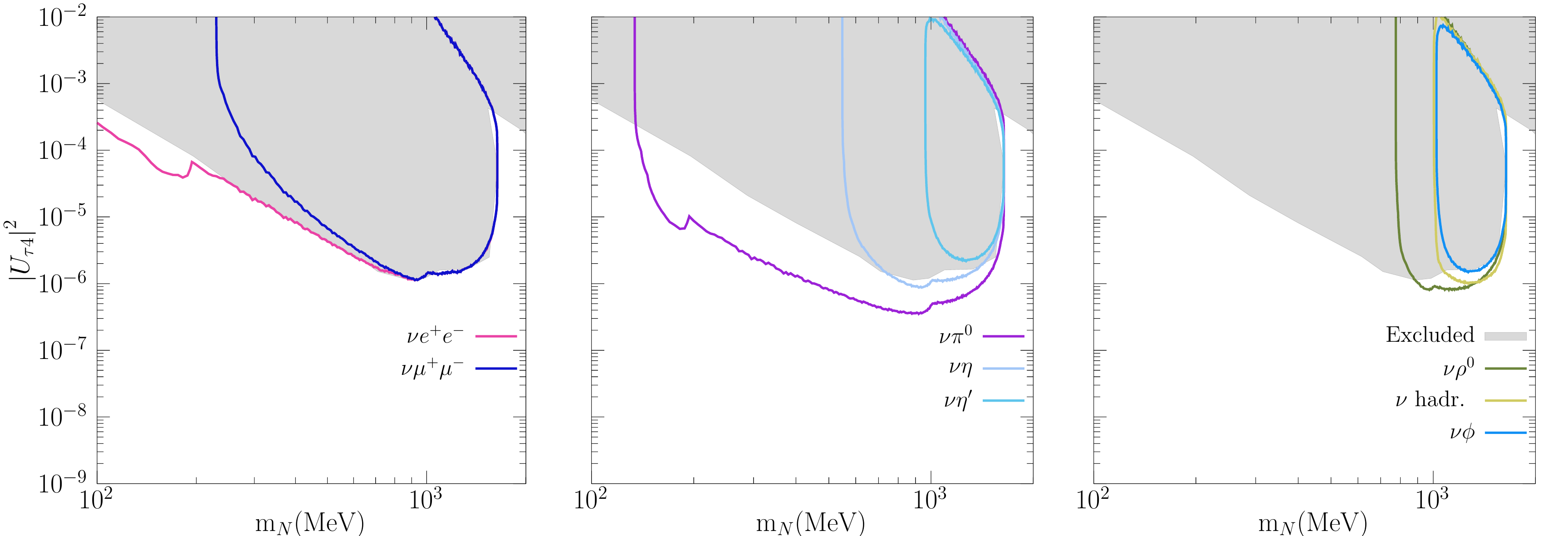}
\end{center}
\caption{\label{fig:HNL_channels} Expected sensitivity of the proposed setup to Heavy Neutral Leptons (HNLs) at $90\%$ C.L., separated by detection channel. In each panel, results
are obtained setting the remaining mixing parameters to zero, assuming backgrounds can be reduced to a negligible level and for perfect detector efficiency. }
\end{figure*}

This Appendix summarizes the expected sensitivities obtained for a HNL decaying into specific final states, so they can be easily recasted once the information on expected efficiencies and background is available for each channel. Our results are shown in Fig.~\ref{fig:HNL_channels}, where the different lines correspond to different decay modes of the HNL.

\bibliography{references.bib}

\end{document}